\documentclass[sigconf]{acmart}
\pdfoutput=1

\usepackage{graphicx}
\usepackage{commath}
\usepackage{mathtools}
\usepackage[braket]{qcircuit}
\usepackage{tikz}
\usetikzlibrary{patterns}
\usetikzlibrary{arrows}
\usepackage{pgfplots}
\pgfplotsset{compat=1.14}
\usepackage{float}
\usepackage{tkz-graph}
\tikzset{
    VertexStyle/.append style = {inner sep=2pt},
    EdgeStyle/.append style = {->},
}
\usepackage{array}
\newcolumntype{C}[1]{>{\centering\let\newline\\\arraybackslash\hspace{0pt}}m{#1}}

\newcommand{\reviewaddition}[1]{{#1}}

\newcommand\plotscale{0.9}
\newcommand{\plotheight}{0.211}

\newcommand\figscale{0.82}
\AtBeginDocument{%
  \providecommand\BibTeX{{%
    \normalfont B\kern-0.5em{\scshape i\kern-0.25em b}\kern-0.8em\TeX}}}

\copyrightyear{2020}
\acmYear{2020}
\setcopyright{acmcopyright}\acmConference[CF '20]{17th ACM International Conference on Computing Frontiers}{May 11--13, 2020}{Catania, Italy}
\acmBooktitle{17th ACM International Conference on Computing Frontiers (CF '20), May 11--13, 2020, Catania, Italy}
\acmPrice{15.00}
\acmDOI{10.1145/3387902.3392617}
\acmISBN{978-1-4503-7956-4/20/05}

\acmSubmissionID{22}

\begin{document}

%%
%% The "title" command has an optional parameter,
%% allowing the author to define a "short title" to be used in page headers.
\title[Time-Sliced Quantum Circuit Partitioning for Modular Architectures]{Time-Sliced Quantum Circuit Partitioning \\ for Modular Architectures}

%
% The "author" command and its associated commands are used to define
% the authors and their affiliations.
% Of note is the shared affiliation of the first two authors, and the
% "authornote" and "authornotemark" commands
% used to denote shared contribution to the research.
\author{Jonathan M. Baker}
\email{jmbaker@uchicago.edu}
\orcid{0000-0002-0775-8274}
\affiliation{%
 \institution{University of Chicago}
 %\city{Chicago}
 %\state{IL}
}

\author{Casey Duckering}
\email{cduck@uchicago.edu}
\orcid{0000-0002-4656-9644}
\affiliation{%
 \institution{University of Chicago}
 %\city{Chicago}
 %\state{IL}
}

\author{Alexander Hoover}
\email{alex8@uchicago.edu}
\orcid{}
\affiliation{%
 \institution{University of Chicago}
 %\city{Chicago}
 %\state{IL}
}

\author{Frederic T. Chong}
\email{chong@cs.uchicago.edu}
\affiliation{%
 \institution{University of Chicago}
 %\city{Chicago}
 %\state{IL}
}

%\author{Ben Trovato}
%\email{trovato@corporation.com}
%\orcid{1234-5678-9012}
%\author{G.K.M. Tobin}
%\authornotemark[1]
%\email{webmaster@marysville-ohio.com}
%\affiliation{%
%  \institution{Institute for Clarity in Documentation}
%  \streetaddress{P.O. Box 1212}
%  \city{Dublin}
%  \state{Ohio}
%  \postcode{43017-6221}
%}

%
% By default, the full list of authors will be used in the page
% headers. Often, this list is too long, and will overlap
% other information printed in the page headers. This command allows
% the author to define a more concise list
% of authors' names for this purpose.
% \renewcommand{\shortauthors}{Trovato and Tobin, et al.}

\begin{abstract}
Current quantum computer designs will not scale.  To scale beyond small prototypes, quantum architectures will likely adopt a modular approach with clusters of tightly connected quantum bits and sparser connections between clusters.  We exploit this clustering and the statically-known control flow of quantum programs to create tractable partitioning heuristics which map quantum circuits to modular physical machines one time slice at a time. Specifically, we create optimized mappings for each time slice, accounting for the cost to move data from the previous time slice and using a tunable lookahead scheme to reduce the cost to move to future time slices. We compare our approach to a traditional statically-mapped, owner-computes model. Our results show strict improvement over the static mapping baseline. We reduce the non-local communication overhead by 89.8\% in the best case and by 60.9\% on average. Our techniques, unlike many exact solver methods, are computationally tractable. 
\end{abstract}

%%
%% The code below is generated by the tool at http://dl.acm.org/ccs.cfm.
%% Please copy and paste the code instead of the example below.
%%
%\begin{CCSXML}
%<ccs2012>
% <concept>
%  <concept_id>10010520.10010553.10010562</concept_id>
%  <concept_desc>Computer systems organization~Embedded systems</concept_desc>
%  <concept_significance>500</concept_significance>
% </concept>
% <concept>
%  <concept_id>10010520.10010575.10010755</concept_id>
%  <concept_desc>Computer systems organization~Redundancy</concept_desc>
%  <concept_significance>300</concept_significance>
% </concept>
% <concept>
%  <concept_id>10010520.10010553.10010554</concept_id>
%  <concept_desc>Computer systems organization~Robotics</concept_desc>
%  <concept_significance>100</concept_significance>
% </concept>
% <concept>
%  <concept_id>10003033.10003083.10003095</concept_id>
%  <concept_desc>Networks~Network reliability</concept_desc>
%  <concept_significance>100</concept_significance>
% </concept>
%</ccs2012>
%\end{CCSXML}


%\ccsdesc[500]{Computer systems organization~Embedded systems}
%\ccsdesc[300]{Computer systems organization~Redundancy}
%\ccsdesc{Computer systems organization~Robotics}
%\ccsdesc[100]{Networks~Network reliability}

%%
%% Keywords. The author(s) should pick words that accurately describe
%% the work being presented. Separate the keywords with commas.
%\keywords{datasets, neural networks, gaze detection, text tagging}

%% A "teaser" image appears between the author and affiliation
%% information and the body of the document, and typically spans the
%% page.
%\begin{teaserfigure}
%  \includegraphics[width=\textwidth]{sampleteaser}
%  \caption{Seattle Mariners at Spring Training, 2010.}
%  \Description{Enjoying the baseball game from the third-base
%  seats. Ichiro Suzuki preparing to bat.}
%  \label{fig:teaser}
%\end{teaserfigure}

%%
%% This command processes the author and affiliation and title
%% information and builds the first part of the formatted document.
\maketitle

\section{Introduction}  \label{introduction}

Quantum computing aims to provide significant speedup to many problems by taking advantage of quantum mechanical properties such as superposition and entanglement \cite{quantum_ml, quantum_chemistry, quantum_optimization}. Important applications such as Shor's integer factoring algorithm \cite{Shor} and Grover's unordered database search algorithm \cite{Grover} provide potentially exponential and quadratic speedups, respectively.

\reviewaddition{Current quantum hardware of the NISQ era \cite{preskill_nisq}, which has on the order of tens to hundreds of physical qubits, is insufficient to run these important quantum algorithms. Scaling these devices even to a moderate sizes with low error rates has proven extremely challenging. Manufacturers of quantum hardware such as IBM and IonQ have had only limited success in extending the number of physical qubits present on a single contiguous piece of hardware. Issues on these devices such as crosstalk error scaling with the number of qubits or increased difficulty in control will limit the size this single-chip architecture can achieve \cite{bruzewicz2019trapped, brown2016co}}.

\begin{figure}
    \centering
    \scalebox{\figscale}{%
    \includegraphics[width=\columnwidth,keepaspectratio=true]{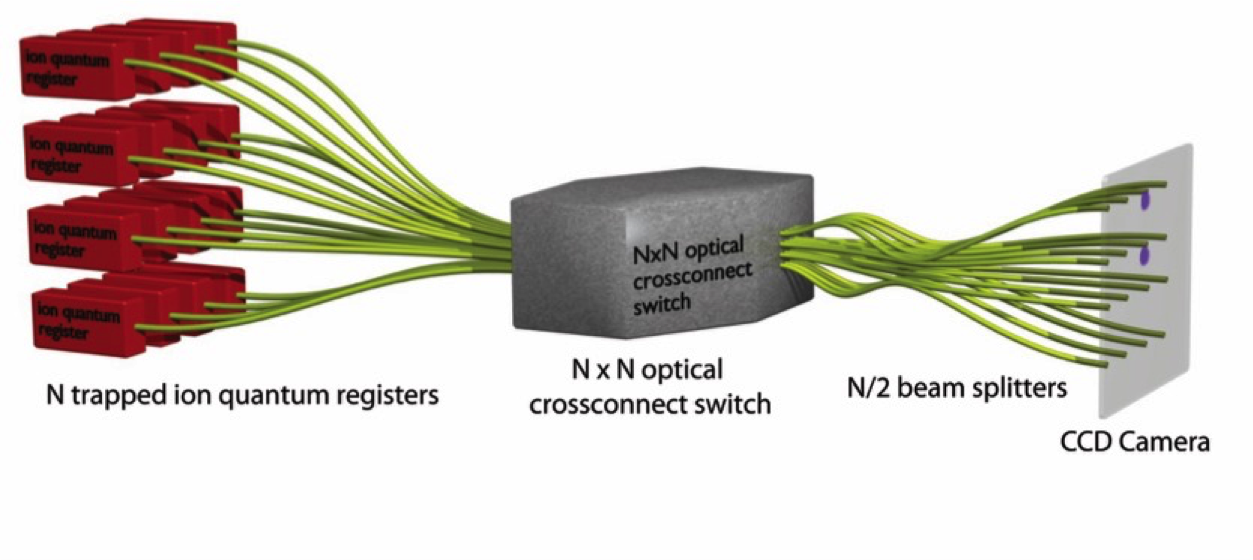}}\vspace*{-.2in}%
    \caption{An example modular architecture of qubits in individual ion traps connected with optics proposed by Monroe et al \cite{modular-ion}.  Communication between traps is supported by photon-mediated entanglement. Similar communication for superconducting qubits \cite{yale-modular} can facilitate modular architectures for that technology.}
    \label{fig:modular-ion}
\end{figure}

\reviewaddition{Due to these challenges, as well as developing technology for communicating between different quantum chips \cite{blakestad2009high, wallraff2018deterministic}, we expect quantum hardware to scale via a modular approach similar to how a classical computer can be scaled increasing the number of processors not just the size of the processors. Two of the leading quantum technologies, ion trap and superconducting physical qubits, are already beginning to explore this avenue and experimentalists project modularity will be the key to moving forward \cite{brecht2016multilayer, devoret2013superconducting, duan2010colloquium, bapat2018unitary, maslov2018outlook, monroe2013scaling, hucul2017spectroscopy}. One such example for ion traps is shown in Figure \ref{fig:modular-ion} where many trapped ion devices are connected via a single central optical switch. Technology such as resonant busses in superconducting hardware or optical communication techniques in ion trap devices will enable a more distributed approach to quantum computing, having many smaller, well-connected devices with sparser and more expensive non-local connections between them. Optimistically, due to current technology in the near term, we expect these non-local communication operations to be somewhere between 5-100x higher latency than in-cluster communication.}

\reviewaddition{With cluster-based approaches becoming more prominent, new compiler techniques for mapping and scheduling of quantum programs are needed. As the size of executable computations increase it becomes more and more critical to employ program mappings exhibiting both adaptivity of dynamic techniques and global optimization of static techniques.  Key to realizing both advantages is to simplify the problem.  Since non-local communication is dominant, we focus on only non-local costs.  This simplification, along with static knowledge of all control flow, allows us to map a program in many timeslices with substantial lookahead for future program behavior. This approach would not be computationally tractable on a non-clustered machine.}  

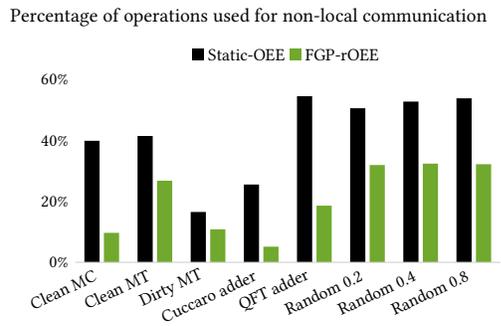
\begin{figure}
    \centering
    \quad\qquad
    \scalebox{\figscale}{%
    % This file was automatically generated by python latextools.
% https://github.com/cduck/latextools
%
% \usepackage{tikz}
% \usepackage{pgfplots}
%
% % Setup package: tikz
% \usetikzlibrary{patterns}
% \usetikzlibrary{arrows}
% \usetikzlibrary{external}
%
% % Setup package: pgfplots
% \pgfplotsset{{compat=1.14}}
%
% Pgfplots figure
\begin{tikzpicture}[baseline,scale=1,trim axis left,trim axis right]
%\pgfplotsset{every axis/.append style={thick}}
\pgfplotsset{every tick label/.append style={font=\small}}
\pgfplotsset{every axis label/.append style={font=\small}}

    \begin{axis}[
        name=plot0,
        title={Percentage of operations used for non-local communication\qquad\qquad\qquad},
        xlabel={},
        ylabel={},
        symbolic x coords={Clean MC,Clean MT,Dirty MT,Cuccaro adder,QFT adder,Random 0.2,Random 0.4,Random 0.8},
        width={\columnwidth},
        height={0.6\columnwidth},
        ybar={2pt},
        bar width={7pt},
        enlargelimits=0.07142857142857142,
        ymin=0, ymax=70.72623934322702,
    %    ytick={0, 25, 50, 75, 100},
    %    ytick distance=25,
        xtick=data,
        ,
        legend style={draw=none, fill=none, at={(0.5,1.03)},anchor=north,font=\small},
        legend columns=-1,
        legend image code/.code={\draw[#1, draw=none] (0em,-0.2em) rectangle (0.6em,0.4em);},
        axis line style={draw=black!20!white},
        axis on top,
        y axis line style={draw=none},
        axis x line*=bottom,
        tick style={draw=none},
        yticklabel={\pgfmathparse{\tick*1}\pgfmathprintnumber{\pgfmathresult}\%},
        clip=false,
        enlarge y limits=0,
        ,
        x tick label style={rotate=25, anchor=east},
        ,
        % From: https://tex.stackexchange.com/questions/449620/editing-label-on-bar-chart
        nodes near coords always on top/.style={
            % a new feature since 1.9: allows to place markers absolutely:
    %        scatter/position=absolute,
            every node near coord/.append style={
    %            at={(axis cs:\pgfkeysvalueof{/data point/x},\pgfkeysvalueof{/data point/y})},
    %            draw,      % <-- for debugging only, to check if placement is correct
                anchor=south,
                rotate=0,
                font=\small,
                inner sep=0.2em,
            },
        },
        nodes near coords always on top,
    ]

        \addplot[
            style={
                color=transparent,
                draw=none,
                fill=black,
                ,
                mark=none,
                ,
            }]
        coordinates {
            (Clean MC, 39.805252286810266)
            (Clean MT, 41.37931034482759)
            (Dirty MT, 16.48936170212766)
            (Cuccaro adder, 25.478927203065133)
            (QFT adder, 54.40479949479002)
            (Random 0.2, 50.46335299073294)
            (Random 0.4, 52.64235968865219)
            (Random 0.8, 53.7169772680729)
        };
        \addlegendentry{Static-OEE~~~~};

        \addplot[
            style={
                color=transparent,
                draw=none,
                fill={rgb,255:red,112;green,169;blue,45},
                ,
                mark=none,
                ,
            }]
        coordinates {
            (Clean MC, 9.654561558901683)
            (Clean MT, 26.724137931034484)
            (Dirty MT, 10.795454545454545)
            (Cuccaro adder, 5.121951219512195)
            (QFT adder, 18.556119571348)
            (Random 0.2, 31.86558516801854)
            (Random 0.4, 32.31850117096019)
            (Random 0.8, 32.132132132132135)
        };
        \addlegendentry{FGP-rOEE};

    \end{axis}

\end{tikzpicture}}
    \caption{Non-local communication overhead in circuits mapped to cluster-based machines. Our new mapping scheme FPG-rOEE provides \reviewaddition{reduces the number of operations added for non-local communication} on all benchmarks.}

    \label{fig:com_costs_results}
\end{figure}

For devices with many modular components mapping quantum programs translates readily to a graph partitioning problem with a goal of minimizing edge crossings between partitions. This approach is standard in many classical applications such as high performance parallel computing, etc. \cite{vlsi_partitioning, classical_partitioning, hpc_graph_partitioning} with the goal of minimizing total latency. Here latency is approximated by the total number of times qubits must be shuttled between different regions of the device. Graph partitioning is known to be hard and heuristics are the dominant approach \cite{fiduccia1982linear, park1995algorithms, kernighan1970efficient, hendrickson1995improved, heuristic1}.

\reviewaddition{While this problem is related to many problems in distributed or parallel computing, there are a few very important distinctions. In a typical quantum program, the control flow is statically known at compile time, meaning all interactions between qubits are known. Furthermore, the no-cloning theorem states we cannot make copies of our data, meaning non-local communication between clusters is \textit{required} to interact data qubits. Finally, any additional non-local operations affect not only latency as they would classically but are directly related to the probability a program will succeed since operations in quantum computing are error prone and therefore reducing non-local communication is especially critical for successful quantum program execution.}

Our primary contribution is the development of a complete system for mapping quantum programs to near-term cluster-based quantum architectures via graph partitioning techniques where qubit interaction in-cluster is relatively free compared to expensive out-of-cluster interaction. Our primary goal is to minimize the communication overhead by reducing the number of low-bandwidth, high-latency operations such as moving qubits which are required in order to execute a given quantum program. Rather than partitioning the circuit once to obtain a generally good global assignment of the qubits to clusters, we find a sequence of assignments, one for each time slice in the circuit. This fine-grained approach is much less studied, especially for this class of architectures. With our techniques, we reduce the total number of non-local communication operations by 89.8\% in the best case and 60.9\% in the average case; Figure \ref{fig:com_costs_results} shows a few examples of circuits compiled statically versus with our methods.

The rest of the paper is organized as follows: \reviewaddition{In Section \ref{sec:background}, we introduce the basics of quantum circuits and graph partitioning.} In Section \ref{sec:mapping}, we introduce our proposed methodology for mapping qubits to the clusters of these modular systems, specifically a method for \textit{fine-grained partitioning}. In Section \ref{sec:lookahead}, we introduce a method for applying lookahead weights to tune what is considered \textit{local} at each time slice and evaluate their effect on non-local communication. In Section \ref{sec:benchmarks}, we introduce the benchmarks we test on and present our explicit toolflow for taking quantum programs to a sequence of mappings \reviewaddition{which guarantee interacting qubits are moved into the same partition before each time slice using non-local communication}. In Section \ref{sec:results}, we present our results and provide a brief discussion, and in Section \ref{sec:prior}, we present a summary of related work for hardware mapping. We conclude in Section \ref{sec:conclusion}.

\section{Background}  \label{sec:background}

\subsection{Quantum Programs and Architectures} \label{quantum-architecture}
The typical fundamental unit of quantum information is the qubit (quantum bit). Unlike classical bits which occupy either 1 or 0 at any given time, quantum bits may exist in a superposition of the two basis states $\ket{0}$ and $\ket{1}$. Qubits are manipulated via quantum gates, operations which are both reversible and preserve a valid probability distribution over the basis states. There is a single irreversible quantum operation called measurement, which transforms the qubit to either $\ket{0}$ or $\ket{1}$ probabilistically. Pairs of qubits are interacted via two-qubit gates, which are generally much more expensive in terms of error rates and latency.

There are a variety of competing styles of quantum systems each with a hardware topology specifying the relative location of the machine's qubits. This topology indicates between which pairs of qubits two-qubit interactions may be performed. 

Typical quantum hardware does not readily support long-range multi-qubit operations but does provide a mechanism for moving qubits, either by swapping qubits (in the case of nearest neighbor or 2D-grid devices), teleportation via photon mediated entanglement, physically moving qubits (as in ion-trap devices), \reviewaddition{or a resonant bus (as in superconducting devices)}. Interacting qubits which are distant generate additional latency which is undesirable for near-term qubits with limited coherence time (the expected lifetime of a qubit before an error). \reviewaddition{These machines have expected error rates on the order of 1 in every 100-1000 two-qubit gates \cite{ionq, ibm_error}, and non-local communication has error on average 10-100x worse.} % \ref{tab:est_cost}

In this paper, we are motivated by a specific set of architectures or extensions to such architectures, as in \cite{schuster_machine, ion1, ion2, ion3}. In these devices, qubits are arranged into several regions of high connectivity \reviewaddition{with expensive communication between the clusters, referred to as non-local communication.} These devices naturally lend themselves to mapping techniques which utilize partitioning algorithms. 

\begin{figure}
    \centering%
    \scalebox{\figscale}{
    \input{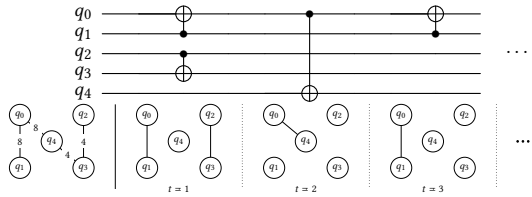}%
    }%
    \\%
    \scalebox{\figscale}{%
    \resizebox{\linewidth}{!}{    \begin{tikzpicture}
        \tikzset{
            VertexStyle/.append style = {
                inner sep=0.2em,
            },
            EdgeStyle/.append style = {
                -,
            }
        }
        
        % Interaction graph
        \Vertex[Math,x=-12em,y=12em]{q_0}
        \Vertex[Math,x=-6em,y=12em]{q_2}
        \Vertex[Math,x=-12em,y=7em]{q_1}
        \Vertex[Math,x=-6em,y=7em]{q_3}
        \Vertex[Math,x=-9em,y=9.5em]{q_4}
        \Edge[label=8](q_0)(q_1)
        \Edge[label=4](q_2)(q_3)
        \Edge[label=8](q_0)(q_4)
        \Edge[label=4](q_3)(q_4)
        
        \draw [solid] (-3em,13em) -- (-3em,5em);
        %%%%%%%%%%%%%%%%%%%%%
        
        \Vertex[Math,x=0em,y=12em]{q_0}
        \Vertex[Math,x=6em,y=12em]{q_2}
        \Vertex[Math,x=0em,y=7em]{q_1}
        \Vertex[Math,x=6em,y=7em]{q_3}
        \Vertex[Math,x=3em,y=9.5em]{q_4}
        \Edge(q_0)(q_1)
        \Edge(q_2)(q_3)
        
        \draw [dotted] (9em,13em) -- (9em,5em);
        
        \Vertex[Math,x=12em,y=12em]{q_0}
        \Vertex[Math,x=18em,y=12em]{q_2}
        \Vertex[Math,x=12em,y=7em]{q_1}
        \Vertex[Math,x=18em,y=7em]{q_3}
        \Vertex[Math,x=15em,y=9.5em]{q_4}
        \Edge(q_0)(q_4)
        
        \draw [dotted] (21em,13em) -- (21em,5em);
        
        \Vertex[Math,x=24em,y=12em]{q_0}
        \Vertex[Math,x=30em,y=12em]{q_2}
        \Vertex[Math,x=24em,y=7em]{q_1}
        \Vertex[Math,x=30em,y=7em]{q_3}
        \Vertex[Math,x=27em,y=9.5em]{q_4}
        \Edge(q_0)(q_1)
        
        \draw [dotted] (33em,13em) -- (33em,5em);
        
        % \Vertex[x=0em,y=0em,  L=q0]{0a}
        % \Vertex[x=0em,y=-3em, L=q1]{1a}
        % \Vertex[x=0em,y=-6em, L=q2]{2a}
        % \Vertex[x=0em,y=-9em, L=q3]{3a}
        % \Vertex[x=0em,y=-12em,L=q4]{4a}
        % \tikzset{EdgeStyle/.append style = {-}}
        % \Edge(0a)(1a)
        % \Edge(2a)(3a)

        % \Vertex[x=4em,y=0em,  L=q0]{0b}
        % \Vertex[x=4em,y=-3em, L=q1]{1b}
        % \Vertex[x=4em,y=-6em, L=q2]{2b}
        % \Vertex[x=4em,y=-9em, L=q3]{3b}
        % \Vertex[x=4em,y=-12em,L=q4]{4b}
        % \tikzset{EdgeStyle/.append style = {-}}
        % \tikzset{EdgeStyle/.append style = {bend left = 25}}
        % \Edge(0b)(4b)

        % \Vertex[x=8em,y=0em,  L=q0]{0c}
        % \Vertex[x=8em,y=-3em, L=q1]{1c}
        % \Vertex[x=8em,y=-6em, L=q2]{2c}
        % \Vertex[x=8em,y=-9em, L=q3]{3c}
        % \Vertex[x=8em,y=-12em,L=q4]{4c}
        % \tikzset{EdgeStyle/.append style = {-}}
        % \tikzset{EdgeStyle/.append style = {bend left = 0}}
        % \Edge(0c)(1c)

        % \Vertex[x=12em,y=0em,  L=q0]{0d}
        % \Vertex[x=12em,y=-3em, L=q1]{1d}
        % \Vertex[x=12em,y=-6em, L=q2]{2d}
        % \Vertex[x=12em,y=-9em, L=q3]{3d}
        % \Vertex[x=12em,y=-12em,L=q4]{4d}
        % \tikzset{EdgeStyle/.append style = {-}}
        % \tikzset{EdgeStyle/.append style = {bend left = 25}}
        % \Edge(0d)(4d)

        % \Vertex[x=16em,y=0em,  L=q0]{0e}
        % \Vertex[x=16em,y=-3em, L=q1]{1e}
        % \Vertex[x=16em,y=-6em, L=q2]{2e}
        % \Vertex[x=16em,y=-9em, L=q3]{3e}
        % \Vertex[x=16em,y=-12em,L=q4]{4e}
        % \tikzset{EdgeStyle/.append style = {-}}
        % \tikzset{EdgeStyle/.append style = {bend left = 0}}
        % \Edge(0e)(1e)
        
        % Labels
        \draw (3em,5em) -- (3em,5em) node[pos=.5] {$t=1$};
        \draw (15em,5em) -- (15em,5em) node[pos=.5] {$t=2$};
        \draw (27em,5em) -- (27em,5em) node[pos=.5] {$t=3$};
        
        \filldraw[black](35em,9.5em)circle(0.075em);
        \filldraw[black](35.5em,9.5em)circle(0.075em);
        \filldraw[black](36em,9.5em)circle(0.075em);
    \end{tikzpicture}}}%
    \caption{(Top) An example of a quantum program with single-qubit gates not shown. The inputs are on the left and time flows to the right toward the outputs. The two-qubit operations here are CNOT (controlled-NOT).
    (Bottom) The graph representations of the quantum circuit of the above circuit. On the far left is the total interaction graph where each edge is weighted by the total number of interactions for the whole circuit. To the right is the sequence of time slice graphs, where an edge is only present if the qubits interact in the time slice. The sum of all time slice graphs is the total interaction graph.}
    \label{fig:sample_program}
\end{figure}

Quantum programs are often represented as circuit diagrams, for example the one in Figure \ref{fig:sample_program}a. We define a \textit{time slice} in a quantum program as a set of operations which are parallel in the circuit representation of the program. We express time slices as a function of both the circuit representation and limitations of the specific architecture. We also define a \textit{time slice range} as a set of contiguous time slices; we also refer to them as \textit{slices} and when no length is specified, it will be assumed to be of length 1. 

For evaluation, we consider two primary metrics: the \textit{width} and the \textit{depth} of a circuit. The width is the total number of qubits used and the depth, or the run time, is the total number of time slices required to execute the program. Qubit movement operations which are inserted in order \reviewaddition{to move interacting qubits into the same partition} contribute to the overall depth of the circuit.

We consider two abstract representations of quantum programs: the total interaction graph and a sequence of time slice interaction graphs, examples of which are found in Figure \ref{fig:sample_program}b. In both representations, each qubit is a vertex and edges between qubits indicate two-qubit operations acting on these qubits. In the total interaction graph, edges are weighted by the total number of interactions between pairs of qubits. In time slice graphs, an edge with weight 1 exists only if the pair of qubits interact at that time slice. 

\subsection{Graph Partitioning}
\subsubsection*{\textbf{Static Partitioning}}
Finding graph partitions is a well studied problem \cite{fiduccia1982linear, park1995algorithms, kernighan1970efficient, hendrickson1995multi} and is used frequently in classical architecture. In this paper, we consider a variant of the problem which fixes the total number of partitions and bounds the total number of elements in each partition. Specifically, given a fixed number of partitions $k$, a maximum partition size $p$, and an undirected weighted graph $G$ with $\abs{V(G)} \le k \cdot p$ we want to find a $k$-way assignment of the vertices to partitions such that the weight of edges between vertices in different partitions is minimized. This can be rephrased in terms of \reviewaddition{statically} mapping a quantum circuit to the aforementioned architectures. Let the total interaction graph be $G$ and let $k$ and $p$ fixed by the topology of the architecture. Minimizing the edge weight between partitions corresponds to minimizing the total number of swaps which must be executed.

Solving for an optimal $k$-way partition is known to be hard \cite{partition_hardness}, but there exist many algorithms which find approximate solutions \cite{kernighan1970efficient, park1995algorithms, fiduccia1982linear}. There are several heuristic solvers such as in \cite{METIS, graph1} which can be used to find approximate $k$-way partition of a graph. However, they often cannot make guarantees about the size of the resulting partitions, preventing us from using them for the fixed size partitioning problem.

\subsubsection*{\textbf{Partitioning Over Time}}
Rather than considering a single graph to be partitioned we instead consider the problem of generating a \textit{sequence} of assignments of qubits to clusters, one for each moment of the circuit. We want to minimize the total number of differences between consecutive assignments, naturally corresponding to minimizing the total number of non-local communications between clusters. This problem is much less explored than the prior approach. Partitioning in this way guarantees interacting qubits will be placed in the same partition making the schedule for the input program immediate. In the case of a static partition, which gives only the initial mapping, a further step is needed to generate a schedule.

\subsubsection*{\textbf{Optimal Compilation and Exact Solvers}}
It is too computationally expensive to find a true optimal solution for even reasonably sized input programs. Use of constraint-based solvers has been used recently to look for optimal and near-optimal solutions \cite{murali, uwsic_spatial_arch1, uwisc_spatial_arch2}. Unfortunately, these approaches will not scale in the near-term let alone to larger, error-corrected devices. We explored the use of these solvers but found them to be too slow. Finding a static mapping with SMT is impractical with more than 30 to 40 qubits, and SMT partitioning over time is impractical when number of qubits times the depth became more than 40.

\section{Mapping Qubits to Clusters}  \label{sec:mapping}
We define an \textit{assignment} as a set of partitions of the qubits, usually at a specific time slice. We present algorithms which take a quantum circuit and output a \textit{path}, defined as a sequence of assignments of the qubits with the condition that every partitioning in the sequence is \textit{valid}. An assignment is valid if each pair of interacting qubits in a time slice are located within the same partition. \reviewaddition{Finally, we define the \textit{non-local communication} between consecutive assignments as the total number of operations which must be executed to transition the system from the first assignment to the second assignment.} The total communication of a path is the sum over all communication along the path. 

\subsection{Computing Non-local Communication}

To compute the non-local communication overhead between consecutive assignments of $n$ qubits, we first construct a directed graph with multiple edges where the nodes in the graph are the partitions and the edges indicate a qubit moving from partition $i$ to partition $j$. We extract all 2-cycles from this graph and remove those edges from the graph. We proceed extracting all 3-cycles, and so on and record the number of $k$-cycles extracted as $c_k$. When there are no cycles remaining, the total number of remaining edges is $r$, and the total communication overhead $C$ is given by
$$C = r + \sum_{k=2}^n (k-1)\cdot c_k$$
The remaining edges indicate a qubit swapping with an unused qubit. We repeat this process for every pair of consecutive assignments in the path to compute the total non-local communication of the path. These cycles specify where qubits will be moved with non-local communication.

\subsection{Baseline Non-local Communication} \label{baseline}
As a baseline we consider using a \textit{Static Mapping} \reviewaddition{using an owner computes model}, which takes into account the full set of qubit interactions for the circuit, providing a generally good assignment of the qubits for the entire duration of the program, called the static assignment. At each time step in the circuit, a good static assignment ensures, on average, qubits are not \textit{too far} from other qubits they will interact with frequently. 
%Therefore, in this scheme when two qubits at some time slice must interact but are not adjacent we swap them together and then back to the static assignment. 
\reviewaddition{We find the assignment which requires the fewest number of swaps from the static assignment but has each pair of interacting qubits in a common partition. \reviewaddition{These assignments form} a path for the computation. We refer to this method of path generation in conjunction with a partitioning algorithm, for example Static Mapping with OEE (Overall Extreme Exchange, discussed further later) is referred to as Static-OEE.}

\subsection{Fine Grained Partitioning}
The primary approach we developed to dynamically map a circuit to hardware is \textit{Fine Grained Partitioning} (FGP). In this algorithm, we find an assignment at every time slice using the time slice graphs. By default, these time slice graphs give only immediately local information about the circuit but have no knowledge about upcoming interactions. Alone, they only specify the constraints of which qubits interact in that time slice. The key advantage for this method is using \textit{lookahead weights}. The main idea is to construct modified time slice graphs capturing more structure in the circuit than the default time slice graphs. We refer to these graphs as time slice graphs with lookahead weights, or \textit{lookahead graphs}.

\begin{figure}
    \centering
    \scalebox{\figscale}{%
    \scalebox{0.8}{\begin{tikzpicture}
        \tikzset{
            VertexStyle/.append style = {
                inner sep=0.2em,
            },
            EdgeStyle/.append style = {
                -,
            }
        }
        % First graph
        \Vertex[Math,x=0em,y=12em]{q_0}
        \Vertex[Math,x=6em,y=12em]{q_2}
        \Vertex[Math,x=0em,y=7em]{q_1}
        \Vertex[Math,x=6em,y=7em]{q_3}
        \Vertex[Math,x=3em,y=9.5em]{q_4}
        \Edge(q_0)(q_1)
        \Edge(q_2)(q_3)
        
        \draw [dotted] (9em,13em) -- (9em,5em);
        
        % Second graph offset 16 extra
        \Vertex[Math,x=12em,y=12em]{q_0}
        \Vertex[Math,x=18em,y=12em]{q_2}
        \Vertex[Math,x=12em,y=7em]{q_1}
        \Vertex[Math,x=18em,y=7em]{q_3}
        \Vertex[Math,x=15em,y=9.5em]{q_4}
        \Edge[label=$\infty+7$](q_0)(q_1)
        \Edge[label=$\infty+3$](q_2)(q_3)
        \Edge[label=$4$](q_3)(q_4)
        %\tikzset{EdgeStyle/.append style={bend left=25}}
        \Edge[label=$8$](q_0)(q_4)

\end{tikzpicture}}}
    \caption{An example of a time slice graph with lookahead weights based on the circuit in Figure \ref{fig:sample_program}. We take the graph from the left and add weight to the edges of qubits that interact in the future. In this case, we take the weight equal to the number of times the qubits will interact in the future.}
    \label{fig:lookahead}
\end{figure}
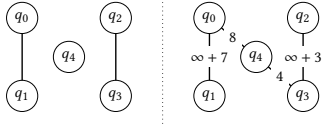

To construct the lookahead graph at time $t$, we begin with the original time slice graph and give the edges present infinite weight. For every pair of qubits we add the weight
$$w_t(q_i,q_j) = \sum_{t< m\le T} I(m,q_i,q_j)\cdot D(m-t)$$
to their edge, where $D$ is some monotonically decreasing, non-negative function, which we call the lookahead function, and $I(m,q_i,q_j)$ is an indicator that is 1 if $q_i$ and $q_j$ interact in time slice $m$ and 0 otherwise, and $T$ is the number of time slices in the circuit. The new time slice graphs consider the remainder of the circuit, more heavily weighting sooner interactions. The effectively infinite weight on edges between interacting qubits is present to guarantee any assignment will place interacting qubits into the same partition.  An example is shown in Figure~\ref{fig:lookahead}.

The final mapping of the qubits in our model is obtained by partitioning each of these time slices. Iteratively, we find the next assignment with a partitioning algorithm, seeded with the assignment obtained from the previous time slice. The first can choose a seed randomly or use the static assignment (presented in \ref{baseline}). The new weights in the time slice graphs will force any movement necessary in the partitioning algorithm. Together, these assignments give us a valid path for the circuit to be mapped into our hardware.

\subsection{Choosing the Partitioning Algorithm}

We assume full connectivity within clusters and the ability to move between clusters. These assumptions give us the liberty to tap into well studied partitioning algorithms. The foundation of many partitioning algorithms is largely considered to be the Kernighan-Lin heuristic for partitioning graphs with bounded partition sizes \cite{kernighan1970efficient, fiduccia1982linear, park1995algorithms}. The KL heuristic selects pairs of vertices in a graph to exchange between partitions based on the weights between the vertices themselves and the total weight between the vertices and the partitions.

We consider a natural extension of the KL algorithm, Overall Extreme Exchange presented by Park and Lee \cite{park1995algorithms}. The OEE algorithm finds a sequence of pairs of vertices to exchange and makes as many exchanges as give it an overall benefit. Using OEE, the Fine Grained Partitioning scheme often over corrects (see Figure \ref{fig:partitioner_results}). If a qubit needs to interact in another partition, then it can ``drag along'' a qubit it is about to interact with because OEE attempts to minimize weight between partitions regardless of its relation to the previous or next time slice graphs. Choosing an optimal partitioning algorithm would not give better solutions to our non-local communication based mapping problem. Instead, we consider a more relaxed version of a partitioning algorithm using the KL heuristic.

\subsubsection*{\textbf{Relaxing the Partitioning Algorithm}}
We provide relaxed version of the algorithm better suited to generating a path over time, called relaxed-OEE (rOEE). We run OEE until the partition is valid for the time slice (all interacting qubits are in the same partition) and then make no more exchanges. This is similar in approach to finding the time slice partitions in our Static Mapping approaches. It is critically important we make our exchange choices using lookahead weights applied to the time slice graphs. Choosing without information about the upcoming circuit provides no insight into which qubits are beneficial to exchange. As a side benefit, making this change strictly speeds up OEE, an already fast heuristic algorithm. Although a strict asymptotic time bound for OEE is difficult to prove, rOEE never took more than a few seconds on any instance it was given.

With such a significant non-local communication overhead improvement (see Figure \ref{fig:partitioner_results}), this relaxed KL partitioning algorithm is much better suited for the problem at hand. It has the ability to take into account local structure in the circuit and avoid over correcting and swapping qubits unnecessarily.

% Section 4: Lookahead weights
    % - candidate functions (expon, constant, norm)
    % - comparison of different scales
    % - choosing a lookahead weight in general
% Should test sigma = 8 as a natural choice for toffoli gates
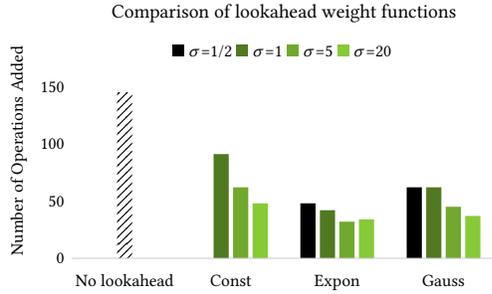
\begin{figure}
    \centering
    \scalebox{\figscale}{%
    % This file was automatically generated by python latextools.
% https://github.com/cduck/latextools
%
% \usepackage{tikz}
% \usepackage{pgfplots}
%
% % Setup package: tikz
% \usetikzlibrary{patterns}
% \usetikzlibrary{arrows}
% \usetikzlibrary{external}
%
% % Setup package: pgfplots
% \pgfplotsset{{compat=1.14}}
%
% Pgfplots figure
\begin{tikzpicture}[baseline,scale=1,trim axis left,trim axis right]
%\pgfplotsset{every axis/.append style={thick}}
\pgfplotsset{every tick label/.append style={font=\small}}
\pgfplotsset{every axis label/.append style={font=\small}}

    \begin{axis}[
        name=plot0,
        title={Comparison of lookahead weight functions},
        xlabel={},
        ylabel={\reviewaddition{Number of Operations Added}},
        symbolic x coords={No lookahead,Const,Expon,Gauss},
        width={\columnwidth},
        height={0.6\columnwidth},
        ybar={2pt},
        bar width={7pt},
        enlargelimits=0.16666666666666666,
        ymin=0, ymax=188.5,
    %    ytick={0, 25, 50, 75, 100},
    %    ytick distance=25,
        xtick=data,
        ,
        legend style={draw=none, fill=none, at={(0.5,1.03)},anchor=north,font=\small},
        legend columns=-1,
        legend image code/.code={\draw[#1, draw=none] (0em,-0.2em) rectangle (0.6em,0.4em);},
        axis line style={draw=black!20!white},
        axis on top,
        y axis line style={draw=none},
        axis x line*=bottom,
        tick style={draw=none},
        yticklabel={\pgfmathparse{\tick*1}\pgfmathprintnumber{\pgfmathresult}},
        clip=false,
        enlarge y limits=0,
        ,
        x tick label style={},
        ,
        % From: https://tex.stackexchange.com/questions/449620/editing-label-on-bar-chart
        nodes near coords always on top/.style={
            % a new feature since 1.9: allows to place markers absolutely:
    %        scatter/position=absolute,
            every node near coord/.append style={
    %            at={(axis cs:\pgfkeysvalueof{/data point/x},\pgfkeysvalueof{/data point/y})},
    %            draw,      % <-- for debugging only, to check if placement is correct
                anchor=south,
                rotate=0,
                font=\small,
                inner sep=0.2em,
            },
        },
        nodes near coords always on top,
    ]

        \addplot[
            style={
                color=transparent,
                draw=none,
                fill=black,
                ,
                mark=none,
                bar shift=-13.5pt,
            }]
        coordinates {
            (No lookahead, 0)
            (Const, 0)
            (Expon, 48)
            (Gauss, 62)
        };
        \addlegendentry{$\sigma$=1/2~~~~};

        \addplot[
            style={
                color=transparent,
                draw=none,
                fill={rgb,255:red,80;green,121;blue,32},
                ,
                mark=none,
                bar shift=-4.5pt,
            }]
        coordinates {
            (No lookahead, 0)
            (Const, 91)
            (Expon, 42)
            (Gauss, 62)
        };
        \addlegendentry{$\sigma$=1~~~~};

        \addplot[
            style={
                color=transparent,
                draw=none,
                fill={rgb,255:red,112;green,169;blue,45},
                ,
                mark=none,
                bar shift=4.5pt,
            }]
        coordinates {
            (No lookahead, 0)
            (Const, 62)
            (Expon, 32)
            (Gauss, 45)
        };
        \addlegendentry{$\sigma$=5~~~~};

        \addplot[
            style={
                color=transparent,
                draw=none,
                fill={rgb,255:red,134;green,202;blue,54},
                ,
                mark=none,
                bar shift=13.5pt,
            }]
        coordinates {
            (No lookahead, 0)
            (Const, 48)
            (Expon, 34)
            (Gauss, 37)
        };
        \addlegendentry{$\sigma$=20~~~~};

        \addplot[
            style={
                color=transparent,
                draw=none,
                fill=black,
                pattern=north east lines,
                mark=none,
                bar shift=0pt,
            }]
        coordinates {
            (No lookahead, 145)
            (Const, 0)
            (Expon, 0)
            (Gauss, 0)
        };

    \end{axis}

\end{tikzpicture}}
    \caption{The effect of different lookahead functions with various $\sigma$ on non-local communication in the Cuccaro adder, a very regular circuit, \reviewaddition{ with 76 data and 24 ancilla qubits} using FGP-rOEE. We see the exponential function outperforms the others for a circuit of highly regular structure.}
    \label{fig:lookahead-bar}
\end{figure}

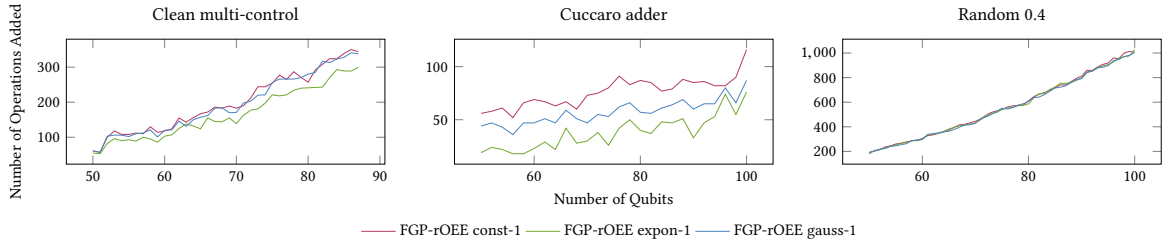
\begin{figure*}
    \centering
    % This file was automatically generated by python latextools.
% https://github.com/cduck/latextools
%
% \usepackage{tikz}
% \usepackage{pgfplots}
%
% % Setup package: tikz
% \usetikzlibrary{patterns}
% \usetikzlibrary{arrows}
% \usetikzlibrary{external}
%
% % Setup package: pgfplots
% \pgfplotsset{{compat=1.14}}
%
\centering
\quad\quad
% Pgfplots figure
\scalebox{\plotscale}{%
\begin{tikzpicture}[baseline,scale=0.85,trim axis left,trim axis right]
%\pgfplotsset{every axis/.append style={thick}}
\pgfplotsset{every tick label/.append style={font=\small}}
\pgfplotsset{every axis label/.append style={font=\small}}

    \begin{axis}[
        name=plot2,
        title={ Clean multi-control },
        xlabel={},
        ylabel={\reviewaddition{Number of Operations Added}},
        width={0.4\textwidth},
        height={\plotheight\textwidth},
    %    xmin=0, xmax=0,
    %    ymin=0,
    %    restrict x to domain=0:0,
    %    xtick distance=25,
        ,
        legend style={
            draw=none,
            at={(0,1)},
            anchor=north west,
            font=\small},
        ,
        clip=false,
    ]

        \addplot[color={rgb,255:red,180;green,56;blue,101}] table[x=x, y=communication-cost-reviewaddition-of-swaps 0, col sep=comma]
            {data/_clean-multi-control-.csv}
        ;%node [anchor=north east] {...};

        \addplot[color={rgb,255:red,112;green,169;blue,45}] table[x=x, y=communication-cost-reviewaddition-of-swaps 1, col sep=comma]
            {data/_clean-multi-control-.csv}
        ;%node [anchor=north east] {...};

        \addplot[color={rgb,255:red,72;green,132;blue,189}] table[x=x, y=communication-cost-reviewaddition-of-swaps 2, col sep=comma]
            {data/_clean-multi-control-.csv}
        ;%node [anchor=north east] {...};
    \end{axis}
\end{tikzpicture}
\quad\quad\quad
% Pgfplots figure
\begin{tikzpicture}[baseline,scale=0.85,trim axis left,trim axis right]
%\pgfplotsset{every axis/.append style={thick}}
\pgfplotsset{every tick label/.append style={font=\small}}
\pgfplotsset{every axis label/.append style={font=\small}}

    \begin{axis}[
        name=plot2,
        title={ Cuccaro adder },
        xlabel={Number of Qubits},
        ylabel={},
        width={0.4\textwidth},
        height={\plotheight\textwidth},
    %    xmin=0, xmax=0,
    %    ymin=0,
    %    restrict x to domain=0:0,
    %    xtick distance=25,
        ,
        legend style={
            draw=none,
            at={(0,1)},
            anchor=north west,
            font=\small},
        ,
        clip=false,
    ]

        \addplot[color={rgb,255:red,180;green,56;blue,101}] table[x=number-of-qubits, y=0, col sep=comma]
            {data/_cuccaro-adder-.csv}
        ;%node [anchor=north east] {...};

        \addplot[color={rgb,255:red,112;green,169;blue,45}] table[x=number-of-qubits, y=1, col sep=comma]
            {data/_cuccaro-adder-.csv}
        ;%node [anchor=north east] {...};

        \addplot[color={rgb,255:red,72;green,132;blue,189}] table[x=number-of-qubits, y=2, col sep=comma]
            {data/_cuccaro-adder-.csv}
        ;%node [anchor=north east] {...};
    \end{axis}
\end{tikzpicture}
\quad\quad\quad
% Pgfplots figure
\begin{tikzpicture}[baseline,scale=0.85,trim axis left,trim axis right]
%\pgfplotsset{every axis/.append style={thick}}
\pgfplotsset{every tick label/.append style={font=\small}}
\pgfplotsset{every axis label/.append style={font=\small}}

    \begin{axis}[
        name=plot2,
        title={ Random 0.4 },
        xlabel={},
        ylabel={},
        width={0.4\textwidth},
        height={\plotheight\textwidth},
    %    xmin=0, xmax=0,
    %    ymin=0,
    %    restrict x to domain=0:0,
    %    xtick distance=25,
        ,
        legend style={
            draw=none,
            at={(0,1)},
            anchor=north west,
            font=\small},
        ,
        clip=false,
    ]

        \addplot[color={rgb,255:red,180;green,56;blue,101}] table[x=x, y=0, col sep=comma]
            {data/_random-04-.csv}
        ;%node [anchor=north east] {...};

        \addplot[color={rgb,255:red,112;green,169;blue,45}] table[x=x, y=1, col sep=comma]
            {data/_random-04-.csv}
        ;%node [anchor=north east] {...};

        \addplot[color={rgb,255:red,72;green,132;blue,189}] table[x=x, y=2, col sep=comma]
            {data/_random-04-.csv}
        ;%node [anchor=north east] {...};
    \end{axis}
\end{tikzpicture}}

\quad\quad
\scalebox{\plotscale}{%
% Pgfplots figure
\begin{tikzpicture}[baseline,scale=0.85,trim axis left,trim axis right]
%\pgfplotsset{every axis/.append style={thick}}
\pgfplotsset{every tick label/.append style={font=\small}}
\pgfplotsset{every axis label/.append style={font=\small}}

    \begin{axis}[
        name=plot2,
        title={  },
        xlabel={},
        ylabel={},
        width={0.8\textwidth},
        height={50pt},
    %    xmin=0, xmax=0,
    %    ymin=0,
    %    restrict x to domain=0:0,
    %    xtick distance=25,
        ,
        legend style={
            draw=none,
            at={(0.5,0)},
            anchor=south,
            font=\small},
        legend columns=-1,
        clip=false,
        axis line style={draw=none},
        tick style={draw=none},
        xticklabels={},
        yticklabels={},
    ]

        \addplot[color={rgb,255:red,180;green,56;blue,101}] table[x=x, y=fgp-roee-const-1  0, col sep=comma]
            {data/__.csv}
        ;%node [anchor=north east] {...};
        \addlegendentry{FGP-rOEE const-1~~~~};

        \addplot[color={rgb,255:red,112;green,169;blue,45}] table[x=x, y=fgp-roee-expon-1  1, col sep=comma]
            {data/__.csv}
        ;%node [anchor=north east] {...};
        \addlegendentry{FGP-rOEE expon-1~~~~};

        \addplot[color={rgb,255:red,72;green,132;blue,189}] table[x=x, y=fgp-roee-norm-1  2, col sep=comma]
            {data/__.csv}
        ;%node [anchor=north east] {...};
        \addlegendentry{FGP-rOEE gauss-1};
    \end{axis}
\end{tikzpicture}}
    \caption{The non-local communication, measured in number of operations between clusters added, for our representative benchmark circuits mapped by each FGP-rOEE using different lookahead functions, each with $\sigma=1$. The x-axis is the number of input/output qubits.  The remainder are used as ancilla for clean multi-control. The exponential function is better on all instances of Clean multi-control and Cuccaro adder, and there is no substantial advantage of one function over the others in the random circuit.}
    \label{fig:lookahead-results}
\end{figure*}

\section{Lookahead Weights}  \label{sec:lookahead}
Finding a suitable lookahead weight function to use in Fine Grained Partitioning is necessary to maximize the benefit gained from choosing our swaps appropriately between time slices. We only require the lookahead function to be monotonically decreasing and non-negative. Throughout this section, we denote our lookahead weight function as $D$.

%\medskip % fix for spacing
\subsection{Natural Candidates}
We explore a few natural candidate weighting functions from the huge space of possible functions. In each of the functions we explore below, we vary a stretching factor or scale $\sigma$ which can be tuned for the given circuit, providing a trade-off between local and global information.

\subsubsection*{\textbf{Constant Function}}
\[ D(n) = \begin{cases} 
      1 & n\leq \sigma \\
      0 & n > \sigma 
   \end{cases}
\]
A constant function captures a fixed amount of local information in the circuit. This is just the number of times the pair of qubits interact in the next $\sigma$ time slices. For $\sigma = 0$, this function corresponds to no lookahead applied.

\subsubsection*{\textbf{Exponential Decay}}
\[ D(n) = 2^{-n/\sigma}
\]
An exponential is a natural way to model a decaying precedence. When $\sigma\le 1$, any interaction will always have a weight at least as high as the sum of interactions after it.

\begin{table*}[]
  \caption{\reviewaddition{A subset of our benchmarks. Clean multi-control has a maximum size of 87. With more than 87 data qubits and fewer than 13 clean ancilla, the depth of the multi-control decomposition is too large to run on these cluster-based machines with predicted error rates.}}
    \centering
    \reviewaddition{
    % This file was automatically generated by python latextools.
% https://github.com/cduck/latextools
%
%
%
\footnotesize\begin{tabular}{c@{\hskip 0.15in} C{.6cm} C{.6cm} C{.6cm} | C{.6cm} C{.6cm} C{.6cm} | C{.6cm} C{.6cm} C{.6cm} | C{.6cm} C{.6cm} C{.6cm}}
    \hline \hline
    {}&\multicolumn{3}{c|}{\textbf{Clean multi-control}}&\multicolumn{3}{c|}{\textbf{Clean multi-target}}&\multicolumn{3}{c|}{\textbf{Dirty multi-target}}&\multicolumn{3}{c}{\textbf{Cuccaro adder}}  \\
    {Data Qubits}&50&76&87&50&76&100&50&76&100&50&76&100 \\ \hline
    Depth&82&265&846&17&22&99&26&34&99&435&669&885 \\
    Two Qubit Op Count (Unmapped)&760&2040&2488&57&85&99&103&157&99&505&778&1030 \\
    Non-local Comm. Ops (Static-OEE)&288&1297&1928&35&60&169&34&31&169&159&243&365 \\
    Non-local Comm. Ops (FGP-rOEE expon-1)&55&218&299&21&31&72&17&19&72&19&42&76 \\
    \hline

    \multicolumn{13}{c}{} \\

    \hline
    {}&\multicolumn{3}{c|}{\textbf{QFT adder}}&\multicolumn{3}{c|}{\textbf{Random 0.2}}&\multicolumn{3}{c|}{\textbf{Random 0.4}}&\multicolumn{3}{c}{\textbf{Random 0.8}} \\
    {Data Qubits}&50&76&100&50&76&100&50&76&100&50&76&100 \\ \hline
    Depth&72&111&147&15&23&30&28&41&54&46&67&86 \\
    Two Qubit Op Count (Unmapped)&625&1444&2500&246&588&995&477&1156&1997&965&2260&3944 \\
    Non-local Comm. Ops (Static-OEE)&512&1144&2542&180&486&863&344&993&1795&682&1944&3462 \\
    Non-local Comm. Ops (FGP-rOEE expon-1)&131&329&541&96&275&498&181&552&1028&386&1070&1964 \\
   \hline \hline
\end{tabular}

    }
  
    \label{tab:benchmarks-table}
\end{table*}

\subsubsection*{\textbf{Gaussian Decay}}
\[ D(n) = e^{-n^2/\sigma^2}
\]
Similar to an exponential, a Gaussian is natural to model decaying precedence with more weight given to local interactions.

%\medskip % fixing spacing
\subsection{Evaluating Lookahead Functions}
To evaluate the choice of lookahead function as well as choice of $\sigma$, we study Fine Grained Partitioning using rOEE with all of the above candidate functions with varying $\sigma$ on benchmarks of various types: those with lots of local structure (a quantum ripple carry adder), those with very little structure (a random circuit), and those which lie somewhere in between (a Generalized Toffoli decomposition).

In Figure \ref{fig:lookahead-bar}, we show an example of a circuit which benefits from having a large scale $\sigma$, the Cuccaro Adder \cite{cuccaro2004adder}. In contrast, all of the random benchmarks benefit from having small $\sigma$ values, functions which decay quickly even for small $n$.

We also compare the different natural lookahead functions we described in the previous section on some representative benchmarks in Figure \ref{fig:lookahead-results}. In these figures, we see the exponential decay has a clear benefit over the rest in the structured circuits of the Multi-Control gate and the Cuccaro Adder. In random circuits, there seems to be no clear benefit to any of the lookahead functions, so long as they have some small lookahead scaling factor. So, we use exponential decay with $\sigma=1$ for our primary benchmarks in Section \ref{sec:benchmarks}.

\section{Experimental Setup}  \label{sec:benchmarks}

All experiments were run on an Intel(R) Xeon(R) Silver 4100 CPU at 2.10 GHz with 128 GB of RAM with 32 cores running Ubuntu 16.04.5. Each test was run on a single core. Our framework runs on Python 3.6.5 using Google's Cirq framework for circuit processing and for implementing our benchmarks \cite{cirq}. For testing exact solvers, we used the Z3 SMT solver \cite{z3}, though results could not be obtained for the size of benchmarks tested because Z3 never completes on problems this size.

\subsection{Benchmarks}

We benchmark the performance of our circuit mapping algorithms on some common sub-circuits used in many algorithms (for example Shor's and Grovers) and, for comparison, on random circuits. Our selection of benchmarks covers a wide variety of internal structure. For every benchmark, we use a representative cluster-based architecture with 100 qubits with 10 clusters each containing 10 qubits but our methods are not limited to any size. We sweep over the number of qubits used from 50 to 100, when in the cases of a few benchmarks the remaining qubits are available for use as either clean or dirty ancilla\footnote{An ancilla is a temporary quantum bit used often to reduce the depth or gate count of a circuit.  ``Clean'' indicates the initial state of the ancilla is known while ``dirty'' means the state is unknown.}. 

\subsubsection*{\textbf{Generalized Toffoli Gate}}
The Generalized Toffoli gate ($C^nU$) is an $n$-controlled $U$ gate for any single qubit unitary $U$ and is well studied \cite{cnx1, cnx2,cnx3,cnx4,cnx5,cnx6}. A $C^nX$ gate works by performing an $X$ gate on the target conditioned on all control qubits being in the $\ket{1}$ state. There are many known decompositions \cite{GidneyBlogPost, He_circuit, Barenco} both with and without the use of ancilla. A complete description of generating these circuits is given by \cite{cnx_decomps}, which provides a method for using clean ancilla.

\subsubsection*{\textbf{Multi-Target Gate}}
The multi-target gate performs a single-qubit gate on many targets conditioned on a single control qubit being in the $\ket{1}$ state. This is useful in several applications such as one quantum adder design \cite{cnx4} and can also be used in the implementation of error correcting codes \cite{ecc}. These circuits can be generated with different numbers of ancilla (both clean and dirty), as given by \cite{cnx_decomps}.

\subsubsection*{\textbf{Arithmetic Circuits}}
Arithmetic circuits in quantum computing are typically used as subcircuits of much larger algorithms like Shor's factoring algorithm and are well studied \cite{cnx3, cnx4, rev_mult}. Many arithmetic circuits, such as modular exponentiation, lie either at the border or beyond the range of NISQ era devices, typically requiring either error correction or large numbers of data ancilla to execute. We examine two types of quantum adders - the Cuccaro Adder and the QFT Adder - as representatives of a class of highly structured and highly regular arithmetic circuits \cite{cuccaro2004adder, qft_adder}.

\subsubsection*{\textbf{Random Circuit}}
The gates presented above have a lot of regular structure when decomposed into circuits. We want to contrast this with circuits with less structure.

We create these random circuits by picking some probability $p$ and some number of samples and generate an interaction between two qubits with probability $p$ for each sample. These circuits have the same structure as QAOA solving a min-cut problem on a random graph with edge probability $p$, so these circuits are a realistic benchmark.

\subsection{Circuit to Hardware}
We begin with a quantum program which is specified at the gate level, consisting of one and two qubit gates. We then generate the total interaction and time slice graphs, where we assume gates are inserted at the earliest possible time. Any further optimization, such as via commutivity or template matching, should be done prior to mapping the program to hardware. We also take the specifications of the hardware, such as number of clusters and the maximum size of the clusters, which constrain possible mappings. 

We use our rOEE as our algorithm for Fine Grained Partitioning. Therefore, we pass the total interaction graph to a static partitioning algorithm to obtain a good starting assignment. This serves as a seed to rOEE rather than starting with a random assignment which may introduce unnecessary starting communication. To the time slice graphs, we apply the lookahead function to obtain the lookahead graphs. We run rOEE on this set of graphs to obtain an assignment sequence such that at every time slice qubits which interact appear in the same bucket. This assignment describes what non-local communication is added before each slice.  Finally, we compute the cost and insert the necessary movement operations into the circuit \reviewaddition{to move interacting qubits into the same partition}, this is a path. As a byproduct, by generating a partitioning over time, we obtain a schedule of operations to be performed.

\section{Results and Discussion}  \label{sec:results}
\begin{figure*}[h!]
    \centering
    \input{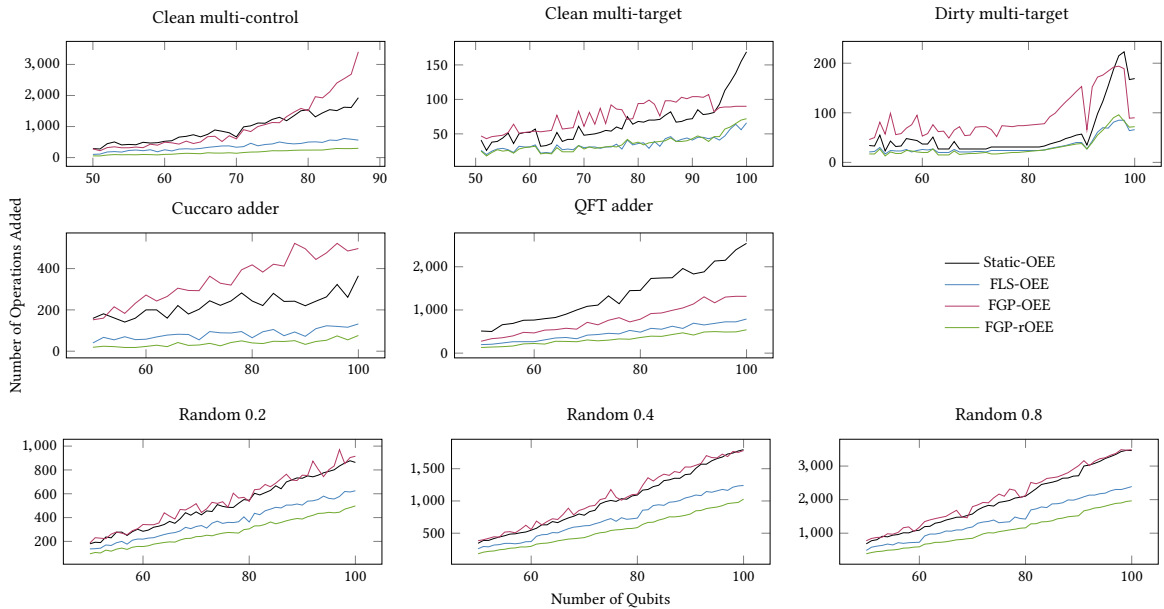}
    \caption{The non-local communication overhead for our benchmark circuits mapped by each mapping algorithm.  The x-axis is the number of qubits that are used in the circuit. \reviewaddition{The y-axis is the number of non-local communication operations inserted to make the circuit executable in our hardware model.} In Clean multi-control, Clean multi-target, and Dirty multi-target, the remainder of the 100 qubits are used as ancilla (clean or dirty determined by the circuit name). FGP-rOEE outperforms all other mapping algorithms on all but the multi-target circuits, and shows substantial improvement over the static baseline. As the size of the circuit increases, rOEE tends to outperform by a greater margin, indicating scales better into the future.}
    \label{fig:partitioner_results}
\end{figure*}

We run our mapping algorithms on each of our benchmark circuits. The results are shown in Figure \ref{fig:partitioner_results}.

Baseline mapping and the original version of OEE perform worse than our best scheme on any benchmark tested. Baseline mapping uses global structure of the graph, but often maintains this structure too much throughout the execution of the circuit. This lack of local awareness and rigid nature of the Static Mapping limits its usefulness. Most out of the box graph partitioning algorithms are designed to only minimize the edge weight between partitions; this will tend to over correct for local structure in the circuit. FGP can overcome this limitation with its choice of partitioning algorithm. By relaxing the partitioning algorithm and not requiring local optimality, we only move qubits until all interacting pairs are together, we require far fewer non-local operations.

%% On Over Correction
The most noticeable changes between FGP-OEE and FGP-rOEE are on the clean multi-control gate with many controls and on the Cuccaro adder. Here, there are often consecutive, overlapping operations with little parallelism. With this structure, after the first operation is performed, the original OEE algorithm will exchange qubits to comply with the next time slice for the next operation. OEE is required to separate qubits which will later interact. To minimize the total crossing weight between partitions, more qubits are shuffled around, usually towards this displaced qubit. In rOEE, this reshuffle optimization never takes place because we terminate once \reviewaddition{each pair of interacting qubits in a time slice is placed in a common partition}. The reshuffling detriments the overall non-local communication when running the circuit because of how often qubits will be displaced from their common interaction partners. In rOEE, not reshuffling keeps the majority of the qubits in sufficiently good spots and the displaced qubit has the opportunity to immediately move back with its interaction partners later.

We include the algorithm Fixed Length Slicing as an alternative not presented in this paper. It is a method with slower computation which explores grouping time slices at fixed intervals. Fixed Length Slicing was consistently the best performing time slice range based mapping algorithm, so we present it in our results. FLS-OEE only beats FGP-rOEE on some instances of the multi-target benchmarks and consistently performs worse on all other benchmarks.

In Figure \ref{fig:com_costs_results}, we show the percentage of operations used for non-local communication for each of the benchmark circuits, and in Table \ref{tab:improvement-table} we show the percent improvement of our algorithm over the baseline. On average, we save over 60\% of the non-local communication operations added. When each non-local communication operation is implemented in hardware, the amount of time each takes is significantly longer than the operations between the qubits in the clusters \cite{mount2016scalable}. Based on current communication technology, we expect these non-local communication operations to take anywhere from 5x to 100x longer than local in-cluster operations. Furthermore, the choice in technology limits how many of these expensive operations can be performed in parallel. 

In Table \ref{tab:est_cost} we compute the estimated running time based on this ratio of costs and show that by substantially reducing the non-local communication via FGP-rOEE, we can drastically reduce the expected run time. We compare our algorithm to the baseline when non-local communication can be performed in parallel (such as in optically connected ion trap devices) and when it is forced to occur sequentially (as when using a resonant bus in superconducting devices). Based on current technology, a 5-10x multiplier is optimistic while 100x is realistic in the near term.

\begin{table}[]
    \caption{Comparing Static-OEE against FGP-rOEE over all benchmarked instances. We obtain improvement across the board with the worst case still reducing non-local communication by 22.6\%.}
    \label{tab:improvement-table}
    \centering
    \begin{tabular}{c@{\hskip 0.3in} c@{\hskip 0.2in} c@{\hskip 0.2in} c}
    \hline \hline 
    \textbf{\% Reduction} & \textbf{min} & \textbf{max} & \textbf{gmean} \\ \hline
    Clean multi-control & 78.1 & 84.9 & 81.9 \\ %\hline
    Clean multi-target & 30.8 & 59.6 & 44.7 \\ %\hline
    Dirty multi-target & 22.6 & 65.1 & 39.9 \\ %\hline
    Cuccaro adder & 79.1 & 89.8 & 85.0 \\ %\hline
    QFT adder & 76.6 & 84.5 & 81.5 \\ %\hline
    Random 0.2 & 52.4 & 57.8 & 55.3 \\ %\hline
    Random 0.4 & 53.6 & 59.0 & 57.0 \\ %\hline
    Random 0.8 & 57.0 & 60.4 & 59.1 \\ \hline
   \textbf{ Aggregate} & \textbf{22.6} & \textbf{89.8} & \textbf{60.9}  \\ \hline \hline
\end{tabular}

\end{table}

\begin{table}[]
    \caption{Estimated execution time of the clean multi-control benchmark with 76 data qubits and 24 ancilla. Two-qubit gates take 300ns \cite{ibm_error} and the multiplier indicates how many times longer non-local communication operations take.}
    \label{tab:est_cost}
    \centering
    % \small\begin{tabular}{c|rr|rr} \hline\hline
%     {} & \multicolumn{2}{c|}{\textbf{Sequential Comm.}} &
%     \multicolumn{2}{c}{\textbf{Parallel Comm.}} \\
%     {Multiplier} & Static-OEE & FGP-rOEE & Static-OEE & FGP-rOEE \\\hline
%     $5x$ & 2025.0 $\mu$s & 406.5 $\mu$s & 673.5 $\mu$s & 256.5 $\mu$s\\
%     $10x$ &  3970.5 $\mu$s & 733.5 $\mu$s & 1267.5 $\mu$s & 433.5 $\mu$s\\
%     $100x$ & 38989.5 $\mu$s & 6619.5 $\mu$s & 11959.5 $\mu$s & 3619.5 $\mu$s\\\hline\hline
% \end{tabular}

\footnotesize\begin{tabular}{c|rr|rr} \hline\hline
    {} & \multicolumn{2}{c|}{\textbf{Sequential Comm.}} &
    \multicolumn{2}{c}{\textbf{Parallel Comm.}} \\
    {Multiplier} & Static-OEE & FGP-rOEE & Static-OEE & FGP-rOEE \\\hline
    $5x$ & 2.0 ms & 0.41 ms  & 0.67 ms & 0.26 ms \\
    $10x$ &  4.0 ms & 0.73 ms & 1.3 ms & 0.43 ms\\
    $100x$ & 39 ms & 6.6 ms & 12 ms & 3.6 ms \\\hline\hline
\end{tabular}
\end{table}

\section{Related Work} \label{sec:prior}
Current quantum hardware is extremely restricted and has prompted a great deal of research aimed at making the most of current hardware conditions. This usually amounts to a few main categories of optimization. The first is circuit optimization at a high level to reduce the number of gates or depth via template matching as in \cite{rw1-template-matching, rw-template-rewriting} or via other optimization techniques as in \cite{optimization-qiskit, automated_optimization}. Other work focuses on optimization at the device level, such as by breaking the circuit model altogether as in \cite{YunongPaper} or by simply improving pulses via Quantum Optimal Control \cite{qoc}. 

At an architectural level, optimization has been studied for many different types hardware with various topologies. The general strategy in most of these works is to reduce SWAP counts with the same motivation as this work, as in \cite{intel1, ai1, siraichi, optimization-qiskit, paler1, paler2, automatic_layout}. Much of this work focuses primarily on linear nearest neighbor (LNN) architectures or 2D lattice architectures as in \cite{lnn1, lnn2, lnn3, lnn4, 2d1}. Some work has focused on ion trap mappings as in \cite{ion_trap_mapping1} though the architecture of this style of device resembles more closely that of a 2D architecture. Some work has recently focused on optimization around specific error rates in near term machines as in \cite{murali, li-ding-xie}. Many of these techniques promise an extension to arbitrary topologies but are not specifically designed to accommodate cluster-based architectures. Work by \cite{qc_paritioning} has explored using graph partitioning to reduce swap counts in near term machines, but their focus is on LNN architectures exclusively. Other work focuses on architectures of the more distant future, namely those with error correction such as in \cite{future1, future2, future3}.

\section{Conclusion}  \label{sec:conclusion}

Alternative to using \reviewaddition{near-optimal} graph partitioning algorithms to find a single static assignment for an entire circuit, we show considering the locality in a circuit during a mapping gives a reduction in the total non-local communication required when running a quantum circuit. There is a natural restriction in using static mappings suggesting the problem of mapping qubits to cluster-based architectures has a different structure than partitioning a single graph for minimum weight between the partitions. Our modification to OEE no longer attempts to optimize the weights at every time slice. It is much more effective in practice to guide the partitioning based on heuristics and not to find the optimal value for every time slice. Optimality at every time slice does not correspond to a global reduction in non-local communication overhead.

% Future work stuff
We propose to use similar schemes for other cluster-based quantum hardware, especially those based on internally connected clusters. In our model, the different clusters of the architecture are also very well connected, but is not limited to only this specific instance of a clustered architecture.

\reviewaddition{Our proposed algorithm produces partitions based on a simplifying assumption about the connectivity of the clusters because the cost of non-local communication is substantially more expensive than any in-cluster operations. Our method can be adapted to other cluster-based architectures by first applying our partitioning algorithm to obtain good clusters of operations and then adding a device-specific scheduling algorithm for scheduling much cheaper in-cluster operations.} 

A relaxed version with well chosen lookahead functions of a heuristic outperforms a well selected initial static mapping. Using lookahead weights has been explored previously, as in \cite{paler1}, and more can be done to better choose the lookahead function, for example based on a metric of circuit regularity. Techniques for mapping which attempt to solve for near optimal mappings will not scale and instead heuristics will be the dominant approach. Our approach is computationally tractable and adaptable to changes in machine architecture, such as additional or varied size clusters. 

Non-local communication overhead in quantum programs makes up a large portion of all operations performed, therefore, minimizing non-local communication is critical. In recent hardware \cite{mount2016scalable}, the cost of moving between clusters makes non-trivial computation impossible with current standards for mapping qubits to hardware. Reducing this hardware bottleneck or finding algorithms to reduce the non-local communication are critical for quantum computation. We reduce this cost substantially in cluster-based architectures \reviewaddition{(see Table \ref{tab:est_cost})}.

%%
%% The acknowledgments section is defined using the "acks" environment
%% (and NOT an unnumbered section). This ensures the proper
%% identification of the section in the article metadata, and the
%% consistent spelling of the heading.
\begin{acks}
This work is funded in part by EPiQC, an NSF Expedition in Computing, under grants CCF-1730449/1832377; in part by STAQ under grant NSF Phy-1818914; and in part by DOE grants DE-SC0020289
and DE-SC0020331.
\end{acks}

%%
%% The next two lines define the bibliography style to be used, and
%% the bibliography file.
\bibliographystyle{ACM-Reference-Format}
\bibliography{references}

\end{document}